\documentclass[12pt,english,prd,amsmath,amssymb]{article}
\usepackage[latin9]{inputenc}
\usepackage[a4paper]{geometry}
\usepackage{array}
\usepackage{multirow}
\usepackage{amstext}
\usepackage{graphicx}

\makeatletter

\newcommand*\LyXThinSpace{\,\hspace{0pt}}
\providecommand{\tabularnewline}{\\}

\usepackage{cite}
\usepackage{graphicx}
\usepackage{dcolumn}
\usepackage{bm}

\@ifundefined{showcaptionsetup}{}{%
 \PassOptionsToPackage{caption=false}{subfig}}
\usepackage{subfig}
\makeatother

\usepackage{babel}
\begin{document}
\begin{center}
\textbf{\Large{}Purity Calculation Method for Events Samples with
Two Identical Particles}
\par\end{center}{\Large \par}

\bigskip{}

\bigskip{}

\begin{center}
{\large{}Valentin Kuzmin}
\par\end{center}{\large \par}

\begin{center}
Moscow State University, Moscow, Russia
\par\end{center}

\bigskip{}

\vspace{3cm}

\begin{center}
\textbf{\large{}Abstract}
\par\end{center}{\large \par}

\medskip{}

This paper studies a method of a two dimensional background calculation
for an analysis of events with two particles of the same type registered
in experiments in high-energy physics. The standard two-dimensional
integration is replaced by an approximation of a specially constructed
one-dimensional function. The number of the signal events is found
by the subtraction of the background events. It allows calculating
the purity of the selection. The procedure does not require a hypothesis
about background and signal shapes. Monte Carlo examples of double
$\textrm{J/}\psi$ samples are used to demonstrate high performance
of the purity calculation method. A comparison of the method with
standard two-dimensional fit of the signal revealed a systematic shift
of the fit results to lower values.

\begin{center}
\bigskip{}
Keywords: data analysis methods; double J/psi; mass spectrum.
\par\end{center}

\newpage{}

\begin{center}
\textbf{\large{}1. Introduction}
\par\end{center}{\large \par}

\medskip{}

If we fit by the Gaussian distribution a one-dimensional (1D) signal
and the need for the reliable result of the n events statistics, then
in the two-dimensional (2D) case one needs for the same fit quality
of the signal to have bigger statistics. The task is complicated by
the fact that the signal shape is not the Gaussian because the accuracy
of the detector reaction depends on the particle kinematics. The additional
uncertainty is the background hypothesis which is much harder to build
for the two-dimensional case. 

The extraction of the cross-section of two similar particles simultaneous
production is a common problem in accelerator experiments. The study
of such processes is complicated by the smallness of the inclusive
cross section of the double production relative to the cross section
of single production of the same particle. For example, the cross
section of double $\textrm{J/}\psi$ \cite{PDG} production is more
than 3 orders less than that of the single $\textrm{J/}\psi$ production
\cite{berezhnoi}. The opportunity to observe pairs with high invariant
masses decreases exponentially with the pair mass. Usually, the main
background at selection has a combinatorial origin. We differ 2 types
of event selection: 1D and 2D cases when we require one or two similar
particles in an event. If we use the Gaussian fit, 2D case requires
much higher statistics than 1D case for the same fit quality. An additional
complication is the fact that the signal shape could not be the Gaussian
because of the dependence of the uncertainty of the detector response
on the particle kinematics. Below, we use the example of the selection
of events with paired $\textrm{J/}\psi$, to describe the method of
the purity calculation in the events with the pair of particles production.

\begin{center}
\medskip{}

\par\end{center}

\medskip{}

\begin{center}
\textbf{\large{}2. Pair selection}\medskip{}

\par\end{center}

In order not to use an abstract description of the method, we use
an example of the $\textrm{ J/}\psi$ pair selection by high-energy
physics detectors to demonstrate the method. Usually, they register
decays of $\textrm{ J/}\psi$ in two muons of opposite charges($\textrm{ J/}\psi\rightarrow\mu^{+}+\mu^{-}$)
\cite{key-1}-\cite{2jpsiD0}. Both muon come from the same vertex.
The invariant mass of the muon pair is $M_{J/\psi}$. We describe
the distribution of the $M_{J/\psi}$ value by the probability distribution
function (PDF)

\begin{equation}
f_{1D}(M_{J/\psi})\,.\label{eq:1dpdf}
\end{equation}

Fig.\ref{fig:1D-pdf} illustrates the distribution which looks like
this function in the $2-5\,Gev/c^{2}$ mass interval. We constructed
the function in Fig.1 to be similar to that in experiments \cite{key-1}-\cite{2jpsiD0}.
We see $\textrm{ J/}\psi$ and $\psi(2S)$ \cite{PDG} peaks over
the combinatorial background created by random combinations of muons.
The shapes of signals are close to the Gaussian ones which could be
deformed by the detector properties. Fig.\ref{fig:1D-pdf} illustrates
the signals are constructed from two Gaussian functions: the base
one with the mean at the PDG value of the particle mass \cite{PDG}
and an additional one with the mean value below the particle mass.
Under the additional Gaussian distribution resides a few percent of
all signal events. We use a polynomial of third order to describe
the background. In our study, we use the distribution in Fig.\ref{fig:1D-pdf}
varying the proportion of $signal/background$ to simulate artificial
single $\textrm{ J/}\psi$ events in our virtual detector. We select
$\textrm{ J/}\psi$ candidates as events with the invariant mass in
the mass window:

\begin{equation}
M_{1}<M_{J/\psi}<M_{2\,}.\label{eq:deltaM}
\end{equation}

\begin{center}
\begin{figure}[h]

\centering{}\includegraphics[scale=0.5]{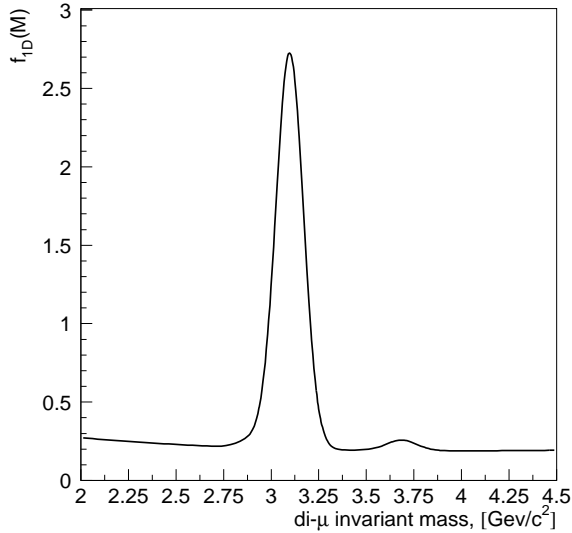}\caption{\label{fig:1D-pdf}1D probability distribution function (\ref{eq:1dpdf}).}
\end{figure}

\par\end{center}

If in some events the detector selects simultaneously two $J/\psi$
candidates, their invariant mass distributions are independent because
we do not differ which particle is the first and which is the second.
The PDF for the events with double $J/\psi$ candidates is 
\begin{equation}
F{}_{2D}(M_{,J/\psi_{1}},M_{,J/\psi_{2}})=f_{1D}(M_{,J/\psi_{1}})f_{1D}(M_{,J/\psi_{2}})\,.\label{eq:2d-pdf}
\end{equation}

We use the purity definition of the sample with $J/\psi$ candidates
as:
\begin{equation}
P=\frac{signal}{signal+background}\,.
\end{equation}

\begin{figure}[h]
\subfloat{\includegraphics[scale=0.5]{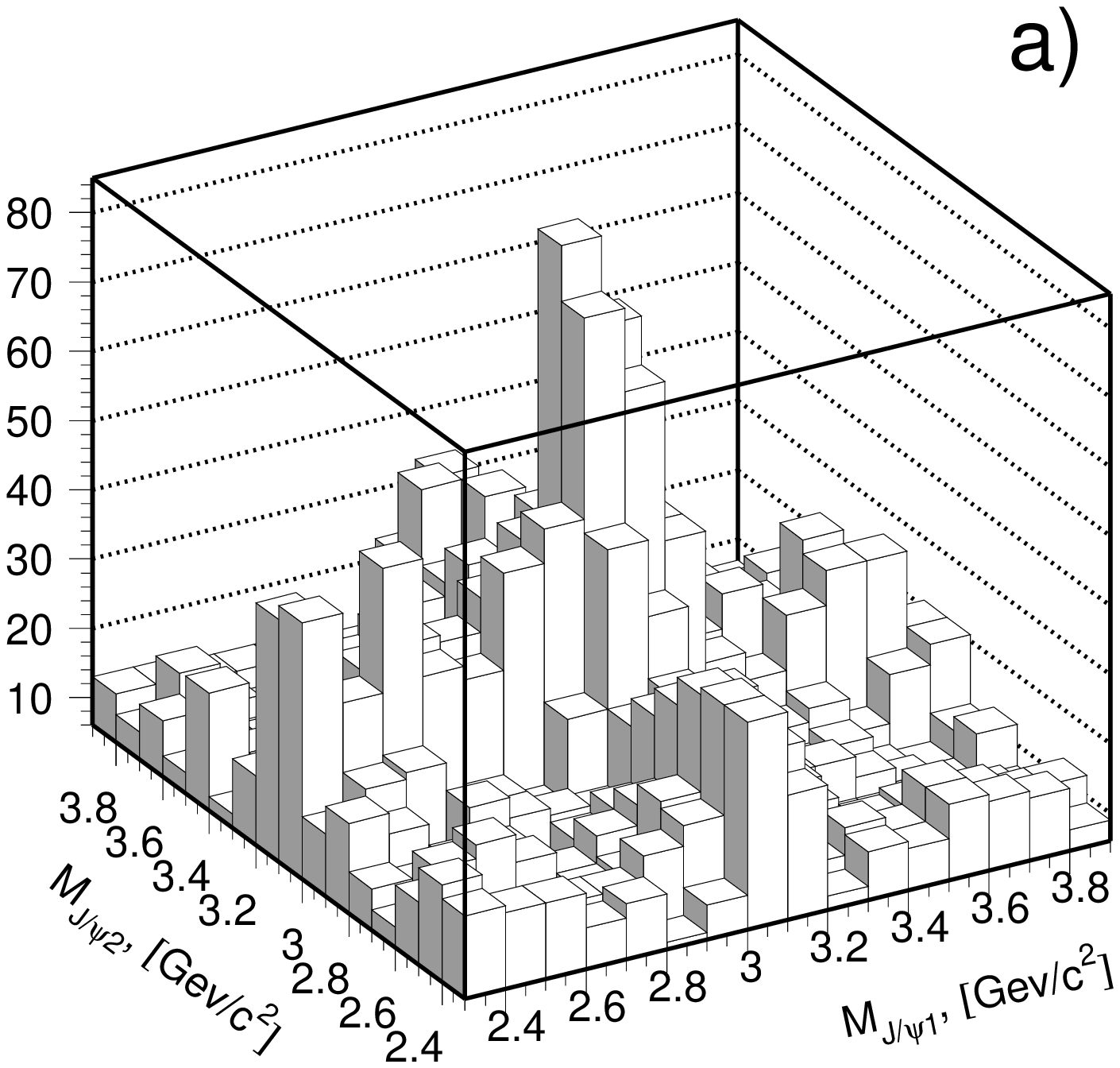}}\subfloat{\includegraphics[scale=0.5]{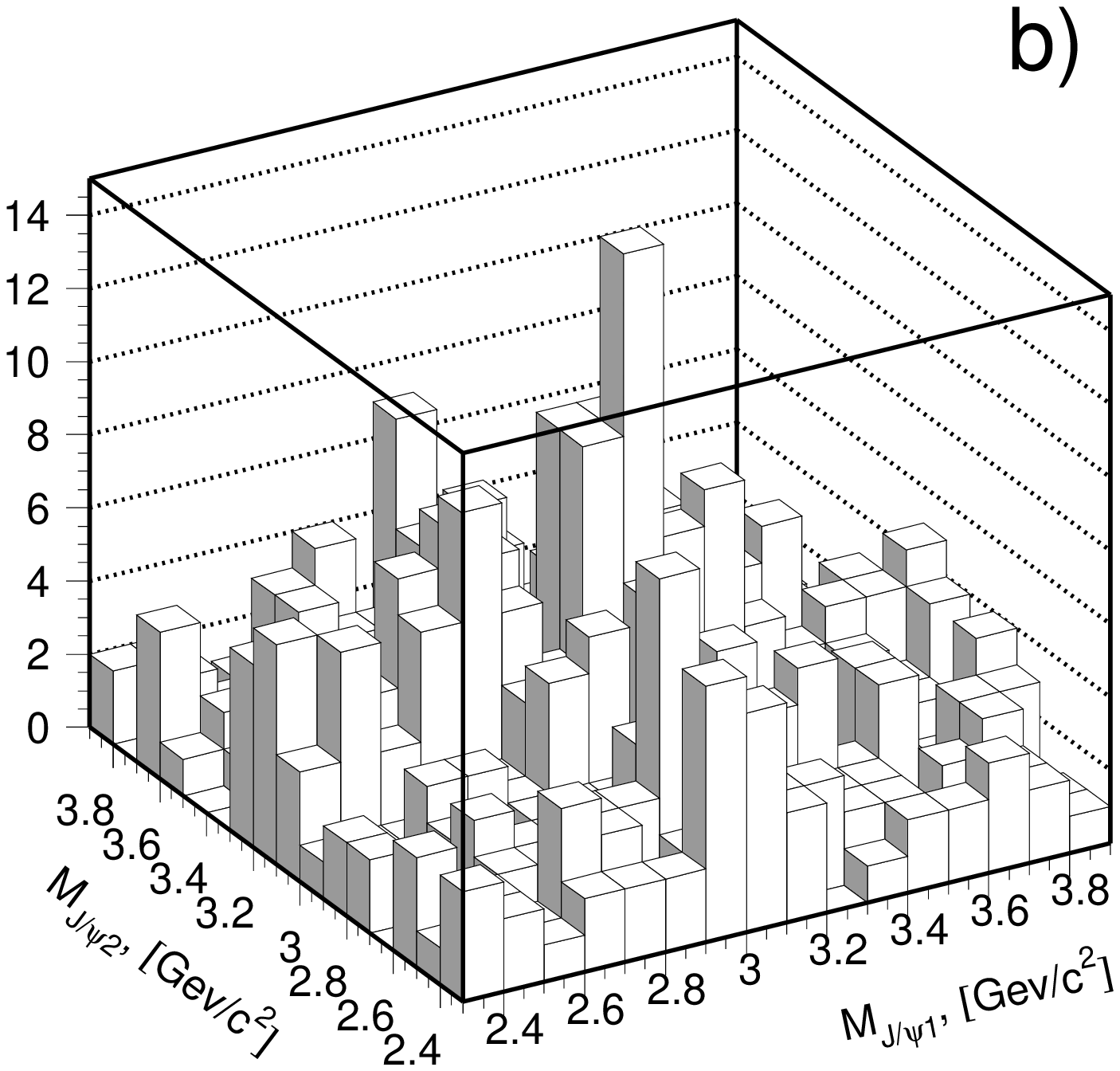}

}\caption{\label{fig:2d_s+g}The scatter plot of $J/\psi$ pairs reconstructed
masses. Presented for the mass window $2.85<M_{J/\psi}<3.3GeV$ are
a) 200 signal events of total 800 events, b) 70 signal events of total
200 events.}

\end{figure}

Fig.\ref{fig:2d_s+g} illustrates two examples of the double $J/\psi$
events registration. Events have been simulated using (\ref{eq:2d-pdf})
and the shape of the distribution in Fig.\ref{fig:1D-pdf}. The cases
differ in terms of the signal amplitude around the $J/\psi$ position
and by the background level. For a real detector, these two examples
correspond to two different event selections: loose (we do not suppress
strongly the combinatorial background) and tight (we suppress the
background and simultaneously decrease the signal). 

The purity of events in Fig.\ref{fig:2d_s+g}a is 25\% whereas the
purity for the sample in Fig.\ref{fig:2d_s+g}b is 35\%. By eye, we
can assume an opposite situation. We need to know the excess of the
double particle events inside the mass window without use of any hypotheses
about signal and background shapes. 

\medskip{}

\medskip{}

\begin{center}
\textbf{\large{}3. Purity of double $J/\psi$ sample }\medskip{}

\par\end{center}

We select events when both muon pairs have the invariant mass inside
a mass window around the $\textrm{J/}\psi$ peak. Such events are
of 3 types: $RR$ (real $\textrm{J/}\psi$ + real $\textrm{J/}\psi$),
$RF$ (real $\textrm{J/}\psi$ + fake $\textrm{J/}\psi$), and $FF$
(fake $\textrm{J/}\psi$ + fake $\textrm{J/}\psi$). Fake $\textrm{J/}\psi$
events come from the combinatorial background. Fig.\ref{fig:2d_s+g}a
illustrates two perpendicular ``mountain ranges'' along the $\textrm{J/}\psi$
mass. They correspond to the background events with one real $\textrm{J/}\psi$
particle and a random muon track combination. In Fig.\ref{fig:2d_s+g}b,
we do not see such phenomenon so evidently. The peak in Fig.\ref{fig:2d_s+g}a
can be due to the superposition of ``mountain ranges.'' In fact, concerning
the form of a two-dimensional background in Fig.\ref{fig:2d_s+g},
we only know that it is a symmetric function around the axis $M_{J/\psi_{1}}=M_{J/\psi_{2}}$.
Anything else can only be our assumptions with unknown accuracy. 

\begin{figure}[h]
\begin{centering}
\includegraphics[scale=0.5]{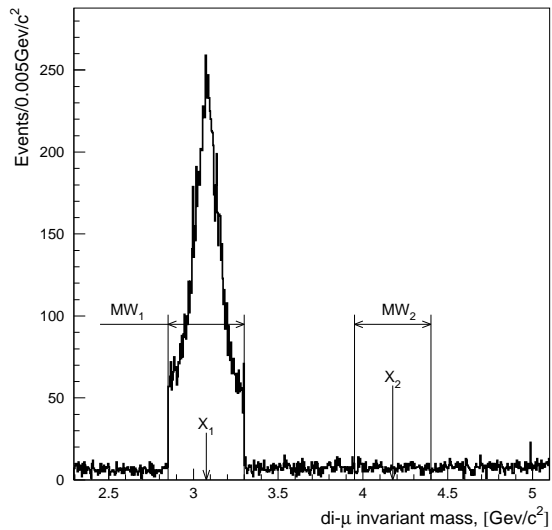}
\par\end{centering}

\caption{\label{example:Di-muon}$J/\psi$ candidate masses in 2D sample. One
candidate has the mass inside the mass window $MW_{1}$, the other
has the mass outside of it. $MW_{2}$ is an arbitrary interval of
the length equal to the length of $WM_{1}$. $X_{1}$ and $X_{2}$
are center masses of $MW_{1}$ and $MW_{2}$.}
\end{figure}

\begin{figure}[h]
\begin{centering}
\includegraphics[scale=0.5]{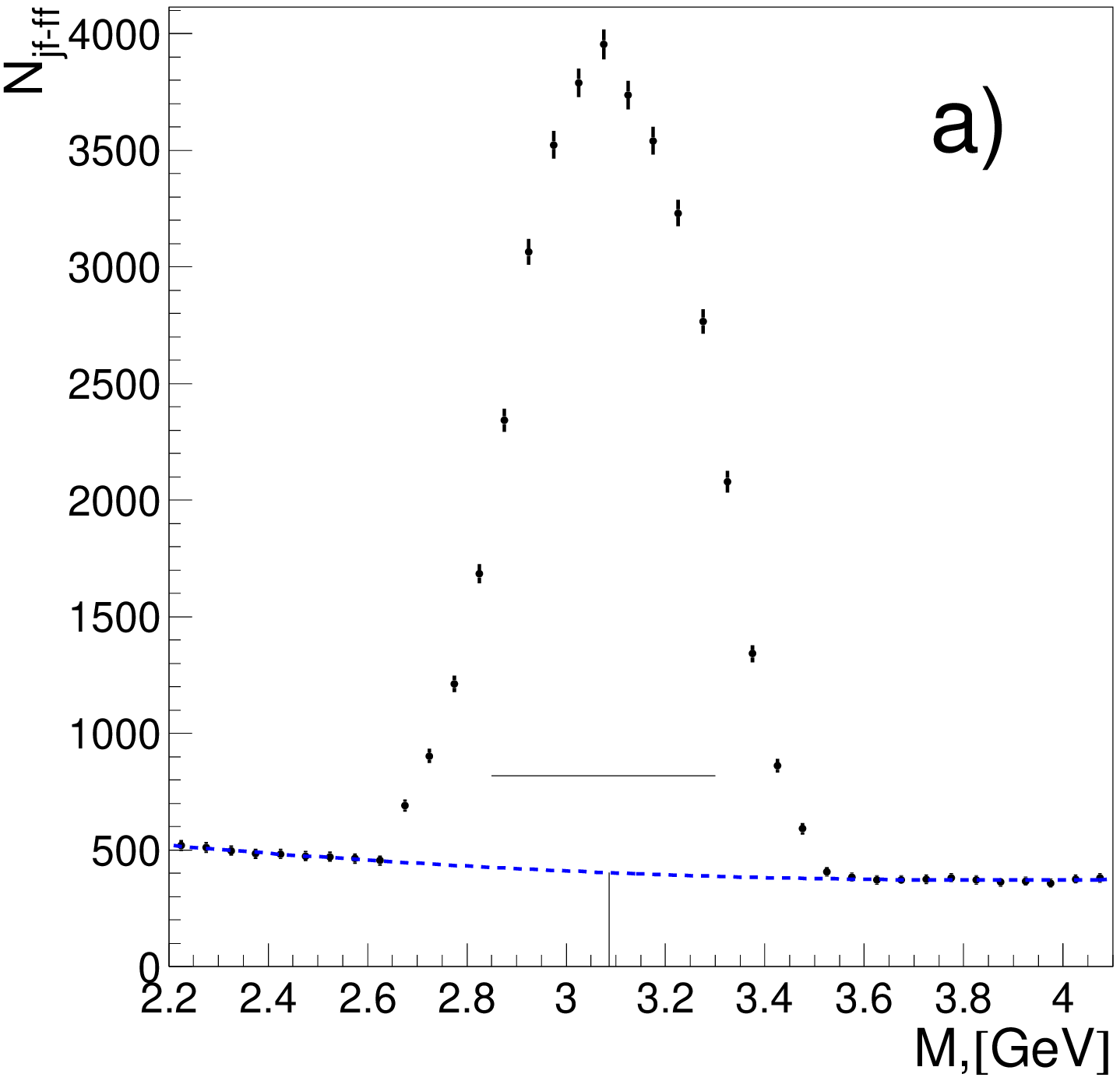}\includegraphics[scale=0.5]{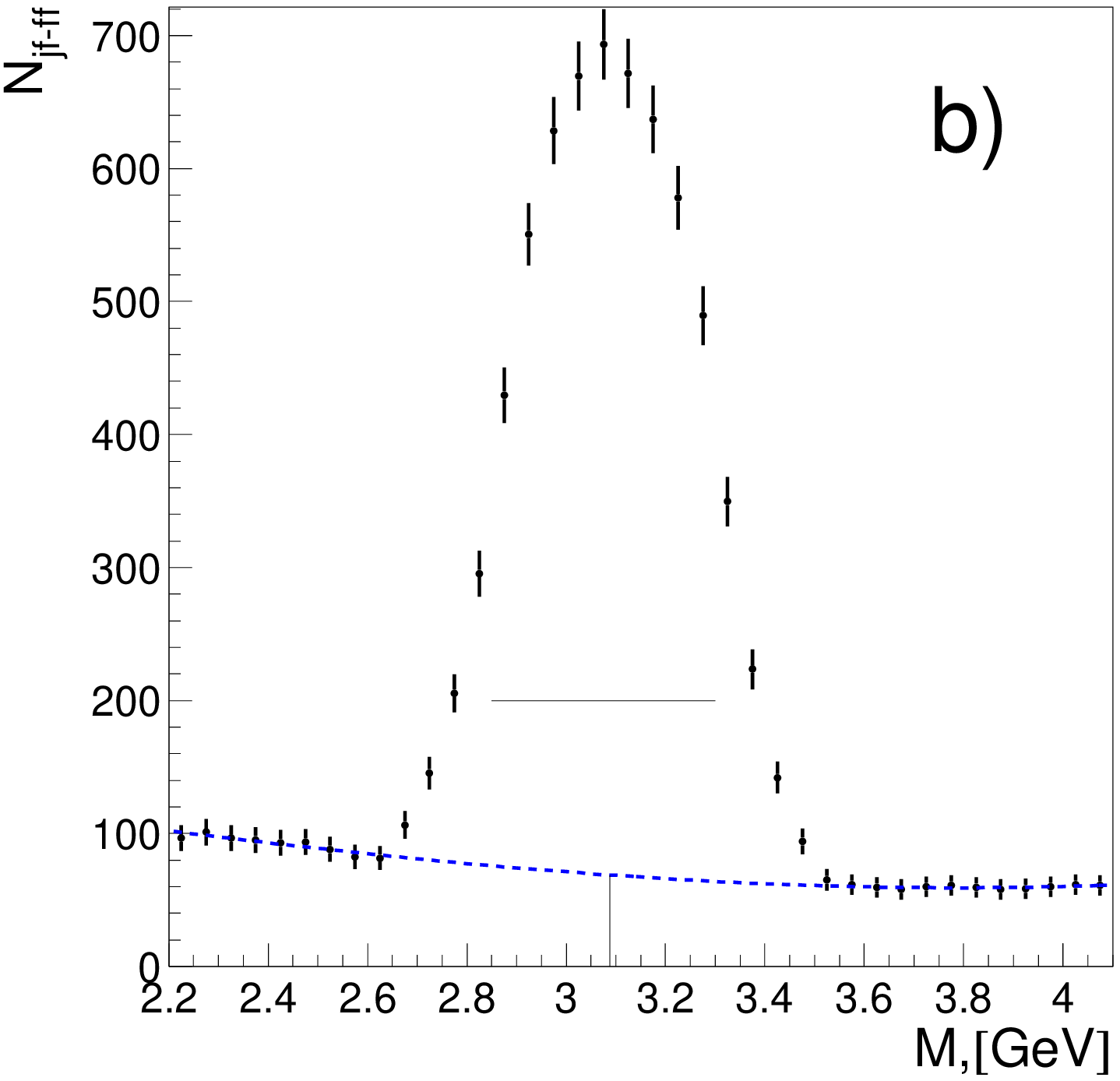}
\par\end{centering}

\caption{\label{N_jfff}$N_{jf-ff}$ and $n_{jf-ff}$ functions. The function
$N_{jf-ff}(M)$ is shown by points. The function $n_{jf-ff}(M)$ is
the dashed line. The horizontal line height is the number of selected
$J/\psi$ candidates in the 2D window. The vertical line shows the
mean value of the invariant mass in $MW_{1}$. a) and b) correspond
to selection cases illustrated in Fig.\ref{fig:2d_s+g} and mass window
$2.85<M_{J/\psi}<3.3\,GeV/c^{2}$.}
\end{figure}

Let us build the distribution of the invariant masses in double particle
events when the invariant mass of one candidate is inside the mass
window $MW_{1}$ (\ref{eq:deltaM}), which we use for double $\textrm{J/}\psi$
selection and the invariant mass of the second candidate is outside
the mass window. Fig.\ref{example:Di-muon} illustrates an example
of the distribution. The distribution defines the function $f_{jf-ff}(M)$
for double particles events with one candidate selected in the mass
window $MW_{1}$. The function $f_{jf-ff}(M)$ outside the mass window
is proportional to the 1D PDF (\ref{eq:1dpdf}) because it is a sum
of independent subsequences of the the random $M_{J/\psi}$ values 

\begin{equation}
f_{jf-ff}(M)=2\int_{WM_{1}}f_{1D}(m)dm\,f_{1D}(M)\,.\label{eq:jfff}
\end{equation}
Using (\ref{eq:jfff}), we build the function 
\begin{equation}
N_{jf-ff}(M)=\int_{WM_{2}}f_{jf-ff}(m)dm\,,
\end{equation}
where $WM_{2}$ is an arbitrary mass interval (see Fig.\ref{example:Di-muon})
and $M$ is its mean value. In practice, it is the sum of events in
the interval $WM_{2}$. If we take $WM_{2}$ far enough from the $\textrm{J/}\psi$
peak, the function $N_{jf-ff}(M)$ provides double number of types
$RF$ and $FF$ events inside the square {[}$WM_{1},WM_{2}]$. Fig.\ref{N_jfff}
illustrates two functions $N_{jf-ff}(M)$ by points. Excluding the
peak region and replacing the function $N_{jf-ff}(M)$ by the fit
of its tails, we get the function $n_{jf-ff}(x)$ illustrated in Fig.\ref{N_jfff}
by the dashed line. Taking into account that the number of random
muon pairs depends on the mass continuously, we obtain that the value
of the function $n_{jf-ff}(M)$ corresponding to the mean value of
$WM_{1}$ is the double number of fake $\textrm{J/}\psi$ pairs of
the 1D projection of all double $\textrm{J/}\psi$ candidates inside
the 2D mass window $WM_{1}$. We use a polynomial to fit tails of
$N_{jf-ff}(M)$. The polynomial is of the degree producing the best
$\chi^{2}/ndf$ value. If $N_{2D}^{all}$ is the number of candidates
inside the 2D mass window $WM_{1}$, the purity of double $\textrm{J/}\psi$
sample in the 2D window can be calculated by the formula: 

\begin{equation}
P_{2D}=1-\left(\frac{n_{jf-ff}(\overline{M}_{J/\psi})}{2N_{2D}^{all}}\right)^{2}\,.\label{eq:pur2d}
\end{equation}
 
The shape of the 2D mass window in used examples is a square. The
described above method allows calculating the purity of 2D samples
for the case of the selection with the 2D mass window being a circle.
It is only needed to use the weight $w$ and the $WM_{1}$ mass window
of the $2R_{MW}$ size around the $\textrm{J/}\psi$ massf or the
calculation of the function $f_{jf-ff}(M)$. $R_{MW}$ is the radius
of the circle mass window used to select $\textrm{J/}\psi$. The weight
$w$ is 
\begin{equation}
w=\frac{\sqrt{R_{MW}^{2}-(m_{J/\psi}-M_{J/\psi})^{2}}}{R_{MW}}\,,\label{eq:weight}
\end{equation}
where $m_{J/\psi}$ is the $\textrm{J/}\psi$ mass \cite{PDG} and
$M_{J/\psi}$ is the mass of the $\textrm{J/}\psi$ candidate which
is inside the mass window $WM_{1}$.

\medskip{}

\medskip{}

\begin{center}
\textbf{\large{}4. Checks using Monte Carlo methods }\medskip{}

\par\end{center}

The contamination of the background in a $\textrm{J/}\psi$ pair sample
increases as the square of the mass window size because the background
is approximately uniform. The signal grows much slower because of
its peaked shape. This means that the $signal/background$ ratio depends
stronger on the size of the mass window as compared to the single
particle selection. The knowledge of the sample purity can help find
the golden mean between the minimal background and good signal. We
performed MC studies using a generic simulation of the detector response
to check results of the previous section. Using (\ref{eq:1dpdf}),
we generated the statistics ($\sim1000$) of 2D MC samples defined
by the mean purity and the event number inside the base mass window
$2.85<M{}_{J/\psi}<3.3\,GeV/c^{2}$. Simulations are in the range
of $1.5<M{}_{J/\psi}<5\,GeV/c^{2}$. To each sample, the described
above method has been applied. Table \ref{Tab_pur_exp} illustrates
purities and signal values found for the two selections $a)$ and
$b)$ in Fig.\ref{fig:2d_s+g} and different mass windows. We give
the average and root mean square (RMS) values. The data confirm the
sensitivity of the suggested method to the purity change and show
high agreement between MC and reconstructed values. 

\begin{center}
\begin{table}[h]
\caption{\label{Tab_pur_exp}Purity of double $\textrm{J/}\psi$ different
selection samples.}

\centering{}%
\begin{tabular}{|>{\raggedright}m{1.3cm}|>{\centering}b{2.3cm}|>{\centering}b{2.4cm}|>{\centering}b{2.4cm}|>{\centering}p{2.4cm}|>{\centering}p{2.4cm}|}
\hline 
version & window,  & \multicolumn{2}{c|}{purity, $\%$} & \multicolumn{2}{c|}{signal, $events$}\tabularnewline
\cline{3-6} 
 & $GeV/c^{2}$ & MC & our method & MC & our method\tabularnewline
\hline 
\hline 
\multirow{2}{1.3cm}{$\quad\:\,$a} & $2.85-3.3$ & $24.9\pm1.6$ & $24.0\pm3.1$ & $199\pm16$ & $193\pm29$\tabularnewline
\cline{2-6} 
 & $2.95-3.2$ & $39.9\pm2.5$ & $39.7\pm3.0$ & $159\pm13$ & $153\pm26$\tabularnewline
\hline 
\multirow{2}{1.3cm}{$\quad\:\,$b} & $2.85-3.3$ & $36.6\pm3.4$ & $36.6\pm4.3$ & $70\pm8$ & $70\pm13$\tabularnewline
\cline{2-6} 
 & $2.95-3.2$ & $50.8\pm4.8$ & $51.8\pm6.2$ & $56\pm8$ & $57\pm11$\tabularnewline
\hline 
\end{tabular}
\end{table}

\par\end{center}

A comparison of our results with the standard fitting of the $signal+background$
sum \cite{2jpsiD0} is an important check of our method. In the simulation
of the detector reaction, we use the signal specified by 6 parameters
(parameters of 2 Gaussians) and the background given by 4 parameters
of the polynomial. In the inverse task, we leave only 2 free parameters:
signal and background values. We take the parameters defining their
shapes from MC and fix them in the fit. This corresponds to exact
knowledge about the detector response. The 2D fit is in the square
of $2.3<M{}_{J/\psi}<3.9\,GeV/c^{2}$. We use average values of MC
sample as initial approach for the fit. 

Table \ref{Tab_pur_mc} illustrates values for the mass window $2.85<M{}_{J/\psi}<3.3\,GeV/c^{2}$
reconstructed by both methods. Just like in Table \ref{Tab_pur_exp},
average and RMS values are given. Different signal and background
conditions have been used in the used MC examples. It allows us to
check the reliability of the purity calculation method in a wide region.
Table \ref{Tab_pur_mc} illustrates high performance of the method
provided in all regions of purities and signals. The 2D fit method,
which uses full information about signal and background shapes, produces
better dispersion at low purities. At high purities, we do not observe
an advantage of the 2D fit method. Moreover, it produces strong systematic
shift which reduces real values of the signal. The investigation of
the reason of this phenomenon is beyond the scope of the present study.
We can only say that at high purities, the mean value of $\chi^{2}$
divided by the number of degrees of freedom is significantly less
than one. In contrast to the 2D fit method, the purity calculation
method suggested in this study does not require a hypothesis about
the background and signal shapes to calculate the purity of the 2D
sample. 

\begin{table}[h]
\caption{\label{Tab_pur_mc}Purity of di-$\textrm{J/}\psi$ samples.}

\centering{}%
\begin{tabular}[b]{|>{\raggedright}m{1.3cm}|>{\centering}p{2cm}|>{\centering}p{2cm}|>{\centering}p{2.3cm}|>{\centering}p{2cm}|>{\centering}p{2cm}|>{\centering}p{2.3cm}|}
\hline 
\multirow{2}{1.3cm}{version} & \multicolumn{3}{c|}{purity, $\%$} & \multicolumn{3}{c|}{signal, $events$}\tabularnewline
\cline{2-7} 
 & MC & 2D fit & our method & MC & 2D fit & our method\tabularnewline
\hline 
\hline 
\multirow{6}{1.3cm}{$\quad\:\,$a} & $5.1\pm0.8$ & $4.8\pm0.6$ & $5.1\pm1.8$ & $41\pm7$ & $39\pm5$ & $42\pm15$\tabularnewline
\cline{2-7} 
 & $15.0\pm1.3$ & $14.2\pm1.0$ & $14.9\pm2.7$ & $122\pm12$ & $119\pm10$ & $120\pm24$\tabularnewline
\cline{2-7} 
 & $30.0\pm1.7$ & $30.6\pm1.6$ & $29.3\pm3.3$ & $242\pm16$ & $245\pm18$ & $236\pm32$\tabularnewline
\cline{2-7} 
 & $50.1\pm1.8$ & $51.9\pm1.7$ & $48.3\pm2.0$ & $403\pm20$ & $417\pm25$ & $390\pm28$\tabularnewline
\cline{2-7} 
 & $75.1\pm1.5$ & $68.4\pm6.2$ & $74.1\pm1.4$ & $600\pm23$ & $547\pm54$ & $592\pm27$\tabularnewline
\cline{2-7} 
 & $95.0\pm0.1$ & $90.8\pm4.0$ & $95.2\pm0.1$ & $759\pm18$ & $723\pm51$ & $759\pm18$\tabularnewline
\hline 
\hline 
\multirow{6}{1.3cm}{$\quad\:\,$b} & $5.0\pm0.2$ & $5.5\pm1.5$ & $4.9\pm2.5$ & $11\pm3$ & $11\pm3$ & $10\pm6$\tabularnewline
\cline{2-7} 
 & $15.0\pm2.5$ & $16.1\pm2.5$ & $14.4\pm3.9$ & $31\pm6$ & $32\pm6$ & $30\pm10$\tabularnewline
\cline{2-7} 
 & $30.0\pm3.3$ & $26.6\pm3.0$ & $29.3\pm4.5$ & $61\pm8$ & $53\pm8$ & $59\pm13$\tabularnewline
\cline{2-7} 
 & $50.1\pm3.5$ & $33.4\pm3.4$ & $50.1\pm3.8$ & $101\pm10$ & $67\pm10$ & $101\pm14$\tabularnewline
\cline{2-7} 
 & $74.9\pm3.0$ & $63.9\pm5.2$ & $75.7\pm2.5$ & $150\pm11$ & $128\pm15$ & $151\pm14$\tabularnewline
\cline{2-7} 
 & $95.0\pm1.6$ & $81.7\pm4.0$ & $95.9\pm1.0$ & $188\pm7$ & $163\pm12$ & $189\pm8$\tabularnewline
\hline 
\end{tabular}
\end{table}
\medskip{}
\medskip{}

\begin{center}
\textbf{\large{}5. Conclusion}\medskip{}

\par\end{center}

The purpose of this paper was to present the method of the two dimensional
background calculations for an analysis of events with two particles
of the same type observed by detectors of high-energy physics. The
standard two-dimensional integration is replaced by an approximation
of a specially constructed one-dimensional function. The number of
the signal events is found by the subtraction of the background from
the number of the total selected events. It allows calculating the
purity value of the selected events sample. The procedure does not
require a hypothesis about shapes of the background and signal events.
High performance of the purity calculation method is demonstrated
using generic Monte Carlo examples of double $\textrm{J/}\psi$ samples.
It has been demonstrated that the 2D fit of the signal like two-dimensional
Gaussian by known shape functions of the signal and background could
produce a significant systematic shift. The shift is bigger at low
statistics. For this reason, it is important to check the fit result
by the suggested method. 

\begin{center}
\bigskip{}
\bigskip{}

\par\end{center}

\bigskip{}
\bigskip{}

\begin{center}
\vspace{-0.5cm}

\par\end{center}
\end{document}